\documentclass{Interspeech2024}

\interspeechcameraready

\usepackage[utf8]{inputenc} 
\usepackage[T1]{fontenc}    
\usepackage{hyperref}       
\usepackage{url}            
\usepackage{booktabs}       
\usepackage{amsfonts}       
\usepackage{nicefrac}       
\usepackage{microtype}      
\usepackage{xcolor}         
\usepackage{cite} 
\hypersetup{colorlinks=true}
\usepackage[page]{appendix}
\usepackage{caption}
\usepackage{subcaption}
\usepackage{graphicx}
\usepackage{algorithm}
\usepackage{algpseudocode}
\usepackage{fancybox,framed}
\usepackage{multirow}
\usepackage{tabularx}
\usepackage{soul}
\usepackage{amsmath}
\usepackage{enumitem}
\usepackage{multirow}
\usepackage{diagbox}
\usepackage[export]{adjustbox}
\usepackage{array}

\title{Synth4Kws: Synthesized Speech for User Defined Keyword Spotting in Low Resource Environments}

\name[affiliation={1}]{Pai}{Zhu}
\name[affiliation={1}]{Dhruuv}{Agarwal}
\name[affiliation={1}]{Jacob W.}{Bartel}
\name[affiliation={1}]{Kurt}{Partridge}
\name[affiliation={1}]{Hyun Jin}{Park}
\name[affiliation={1}]{Quan}{Wang}

\address{
  $^1$Google LLC, Mountain View, CA, U.S.A}
\email{\{paizhu,dhruuv,bartel,kep,hjpark,quanw\}@google.com}


\begin{document}

\maketitle

\begin{abstract}
    
    
    One of the challenges in developing a high quality custom keyword spotting (KWS) model is the lengthy and expensive process of collecting training data covering a wide range of languages, phrases and speaking styles. We introduce \emph{Synth4Kws} -- a framework to leverage Text to Speech (TTS) synthesized data for custom KWS in different resource settings. With no real data, we found increasing TTS phrase diversity and utterance sampling monotonically improves model performance, as evaluated by EER and AUC metrics over 11k utterances of the speech command dataset. In low resource settings, with 50k real utterances as a baseline, we found using optimal amounts of TTS data can improve EER by 30.1\% and AUC by 46.7\%. Furthermore, we mix TTS data with varying amounts of real data and interpolate the real data needed to achieve various quality targets. Our experiments are based on English and single word utterances but the findings generalize to i18n languages and other keyword types.
    


\end{abstract}

\section{Introduction}


With the growing demand for voice control and personal devices from a variety of products such as health rings, smart watches, voice control earbuds, and smart phones etc., a high performance keyword spotting model, with low power consumption and memory footprint becomes increasingly important. Traditionally, keyword spotting (KWS) has focused on predefined keywords such as "Siri", "Hey Google", etc.~\cite{chen2014small,alvarez2019end,zhu2023locale}. However, the surge of personalization and customization in these smart devices has fueled the need for a custom keyword spotting solution that allows users to choose their preferred keyword to trigger a device or software. It is also an integral part of a seamless user experience when interacting with AI agents.

High accuracy ASR systems~\cite{prabhavalkar2024extreme} are commonly used for keyword detection but their energy demand and memory footprint make them less suitable for continuous usage on devices such as phones. Recent studies have employed speech classification models to generate speech embeddings from their encoder hidden layers. Embeddings of different utterances are then compared using cosine similarity to determine the degree of match. As an example, Lin et al.~\cite{lin2020training} used a five-layer convolution network as a shared encoder, and trained 125 independent decoders, each classifying over 40 different keywords. The final output of the shared encoder is used as utterance embedding. While this paper also explores the usefulness of TTS data on speech fields, their experiments and evaluations are optimized for utterance classification. Rybakov et al.~\cite{rybakov2020streaming} further explored different model architectures such as DNN, CNN, SVDF~\cite{alvarez2019end}, CRNN, multi head self attention (MHSA), MHSA-RNNs, etc., and greatly improved classification accuracy.

Such classification models are trained to correctly classify the specific keyword in an utterance, but their utterance embeddings fail to distinguish different phrases, as they are not trained to maximize the embedding distance between utterances with different phrases, and vice versa. To directly optimize embedding distances and improve keyword matching quality, other research~\cite{sacchi2019open, chidhambararajan2022efficientword} has adapted triplet loss from its original application in FaceNet~\cite{schroff2015facenet}. In a batch of examples, they sample an anchor utterance, and then sample a positive utterance that has the same phrase as the anchor utterance, and a negative utterance with a different phrase. The loss function is optimized to minimize the embedding distance of the positive utterance pair, and maximize the embedding distance of the negative pair. They used a GRU two-layer model and achieved reasonable accuracy on the WSJ dataset~\cite{paul1992design}.

In contrast to detecting a pre-defined keyword, in which utterances contain a small set of specific keywords, it is very expensive to collect utterances for a wide range of phrases and have a good distribution of different locales, speaker demographics, speaking styles, and prosodies. However, with the recent advancement of Text-to-Speech(TTS) technologies including MAESTRO~\cite{chen2022maestro}, and Virtuoso~\cite{saeki2023virtuoso}, speech utterances can be generated with over a distribution of the above speech characteristics for more than 100 languages~\cite{saeki2024extending}. Notably, TTS has been used in a range of applications such as automatic dubbing~\cite{effendi2022duration} and voice conversion~\cite{zhang2021transfer}, and proven to be a great data augmentation source in low resource environments. Studies have shown that TTS can be leveraged to augment the training data of other speech models, such as speech recognition~\cite{wang2020improving} and speaker recognition~\cite{huang2021synth2aug}.  Researches~\cite{liu2023leveraging,lim2022user} have also applied TTS to keyword spotting tasks. However those work have focused on using TTS data to improve model quality for different product types, rather than comprehensively evaluate the synthetic data to real data ratio and its quality impact in different resource settings.

In this paper, we show how applying different levels of TTS data improve custom keyword spotting quality. Our experiments and analysis address three practical scenarios:
\begin{itemize}
\item \textbf{No available real data}. We aim to understand how different amounts and diversity of TTS data can improve model quality.
\item \textbf{Low/limited real data}. Similarly, we aim to understand how additional TTS data helps model quality when training with limited real data.
\item \textbf{Need planning for real data}. We aim to understand the quantity of real data need to be collected, on top of sufficient TTS data, to achieve various quality targets.
\end{itemize}

\begin{figure*}[!ht]
	\centering
	\includegraphics[width=1\textwidth]{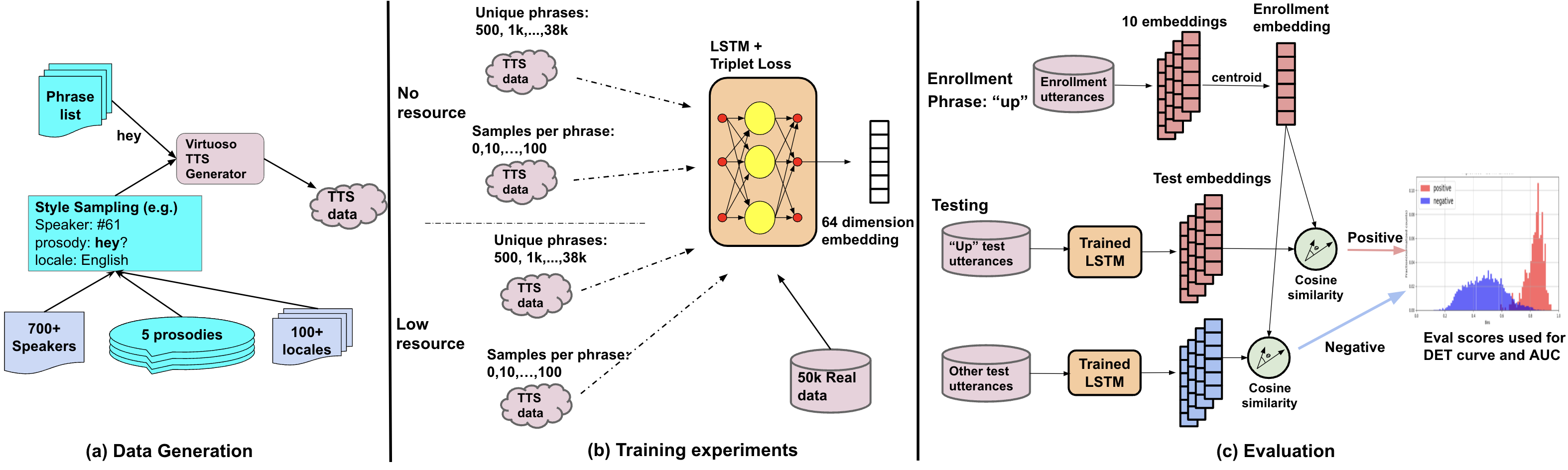}
	\caption{An overview of the Synth4Kws framework: (a) TTS data sampling process; (b) Experiment setup in no-resource and low resource environments with varying phrase diversity and utterance sampling; (c) Evaluation process and metric computation.}
	\label{fig:ab_graph}
\end{figure*}

Our paper is organized as follows. Section~\ref{sec:methods} introduces the the model architectures, an optimized triplet loss batch sampling method used in our experiments, and our TTS model choice with utterance sampling. In Section~\ref{sec:experiment_setups}, we discuss more details about training resources and the TTS experiment setup. We also create a reliable evaluation process and metric definitions to rank model performances. In Section~\ref{sec:exp_results}, we report and analyze our results on the above three scenarios and give our recommendations. Finally, we conclude the paper in Section~\ref{sec:conclusion}.

\section{Methods}
\label{sec:methods}

\subsection{Model Architecture}
\label{ssec:modelarch}
Long Short Term Memory (LSTM)~\cite{lstmref} is a commonly used RNN network that captures long and short term dependencies using memory gates. In this paper we used a standard three LSTM layer model, with hidden layer dimension 384 and output layer dimension 128. This results in a 2.8Mb model (247Kb quantized) that can be continuously run on-device wth reasonable power consumption. We have tried other LSTM model sizes and different architectures including Conformer~\cite{gulati2020conformer}. We choose the 247Kb quantized LSTM model for our TTS experiments based on its good accuracy and low memory footprint, which can fit most device types.

\subsection{Optimized Triplet Loss}
\label{ssec:triplet}
Our triplet loss is optimized based on the traditional method used in Sacchi et al.~\cite{sacchi2019open}. Instead of sampling one positive utterance pair and one negative utterance pair per batch, we use the \textit{generalized-end-to-end approach}~\cite{wan2018generalized} to construct a batch with $X$ phrases and $Y$ sampled utterances per phrase. For each phrase, we used half the sampled utterances as enrollment utterances and the other half as test utterances. Each test utterance embedding is compared to the centroid of the enrollment utterance embeddings. This reduces the variance of enrollment utterance samples and improves training convergence stability. Thus, one batch contains $X*Y/2$ positive examples and $X(X-1)*Y/2$ negative examples.  Since this results in a polarized training distribution with a majority of negative examples and posteriors that can be adversely affected by a skewed training prior, we downweight the negative examples by a factor $\gamma$. In addition, we calculate the loss from all sampled examples using matrix operations to significantly speed up training.

\subsection{TTS utterance sampling}
\label{ssec:tts}
  Virtuoso~\cite{saeki2023virtuoso} is used for TTS data generation for its naturalness and generalization to unseen transcripts and languages. It is built on the foundation of MAESTRO~\cite{chen2022maestro} and leverages different training schemes that combine supervised and unsupervised data (e.g. untranscribed speech and unspoken text data). 
  

As shown in Fig.~\ref{fig:ab_graph}(a), we use the same phrase list containing 38k unique words from MSWC~\cite{mazumder2021multilingual}. For each word, utterances are generated by sampling from 726 speakers and five different prosodies. Virtuoso supports more than 139 locales~\cite{saeki2024extending}. Our experiments focus on English but this approach could benefit other locales even more, especially less common languages with a limited availability of real data.

\section{Data and Experiment Setup}
\label{sec:experiment_setups}
\subsection{Training Resources and Setup}
\label{ssec:trainresource}
Our real utterances are sampled from the MSWC~\cite{mazumder2021multilingual} open source dataset. We process the raw audio into 40 spectral energy features for each 25ms frame. We construct the training batch by randomly choosing eight phrases, and sample 10 utterances per phrase. For each phrase, we use five utterances to build the enrollment centroid. The remaining utterances are used for testing by computing cosine similarity against the enrollment centroids. For training framework, we use Tensorflow/Lingvo~\cite{Shen2019LingvoAM} for its advantages of automatic streaming inference conversion.

\subsection{TTS Experiment Set Up}
\label{ssec:tts_exps}
As mentioned in Section 1, we explore three practical scenarios to analyze TTS's impact on model training and quality. Specifically:
\begin{itemize}
\item \textbf{No real data available}. As shown in the top part of Fig.~\ref{fig:ab_graph}(b), we conduct model training with TTS data alone. To understand the how phrase diversity affects model quality, we tested 500, 1k, 10k and 38k unique phrases respectively with 100 samples per phrase. We also investigate the impact of the number of samples per phrase. With the full phrase diversity, we tested our model with 10, 25, 50, 75, and 100 samples per phrase.

\item \textbf{Limited real data available}. As shown in the bottom part of Fig.~\ref{fig:ab_graph}(b), we start with around 50k real utterances, randomly sampled from MSWC, to build a baseline model for benchmark purposes. Our TTS phrases and utterances sampling methods are the same as the above but sampled utterances are mixed with real data in training. All models are trained from scratch, with different data mixtures.

\item \textbf{Relationship between quality and real data}. As using TTS data only may be inadequate to reach some quality targets, we investigate how much real data is needed. We start with a baseline of 38k unique phrases and 100 samples per phrase of TTS data alone, and incrementally increase the real data amount and analyze the results. We interpolate the model quality vs. real utterance count curves to infer the amount of real data needed for desired quality.
\end{itemize}

\subsection{Eval Dataset and Process}
\label{ssec:eval_process}
Reliable keyword matching evaluation and metrics are important for understanding the effectiveness of TTS data. We use the test split of the commonly used Speech Command Dataset~\cite{warden2018speech} for evaluation. This dataset contains more than 11k utterances from 35 unique phrases. For each phrase, illustrated in Fig.~\ref{fig:ab_graph}(c), we randomly choose 10 utterances as an enrollment set and use the rest for testing. The enrollment and test utterances are treated as a match (positive) if their similarity is above a threshold, and treated as a mismatch (negative) otherwise. By comparing the model prediction and utterances true label (determined by transcripts), we can compute different metrics to rank the models.

\subsection{Eval Metrics}
\label{ssec:eval_metrics}

\begin{figure}[!ht]
	\centering
	\includegraphics[width=4cm, height=4cm]{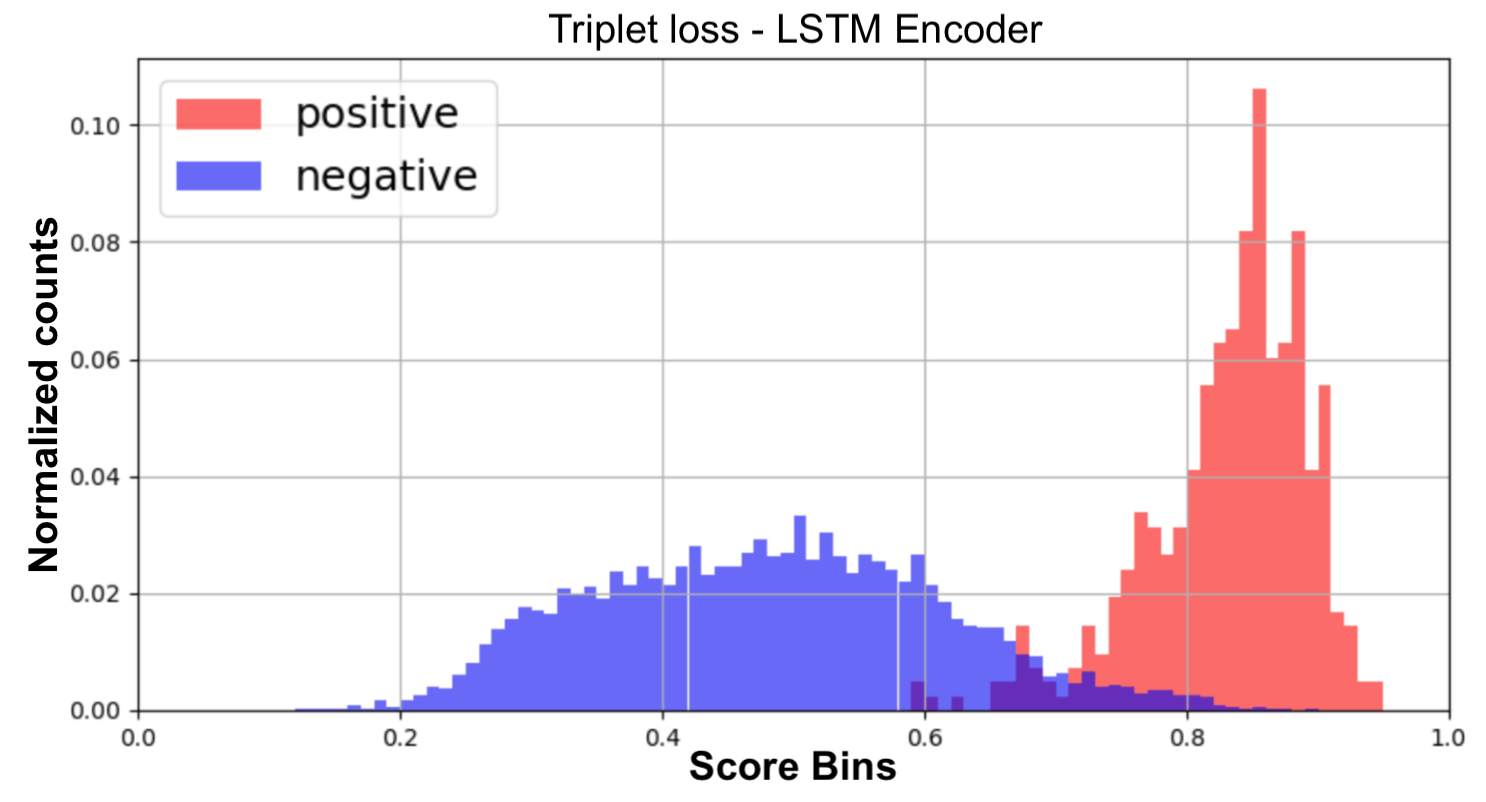}
	\includegraphics[width=3.9cm, height=3.9cm]{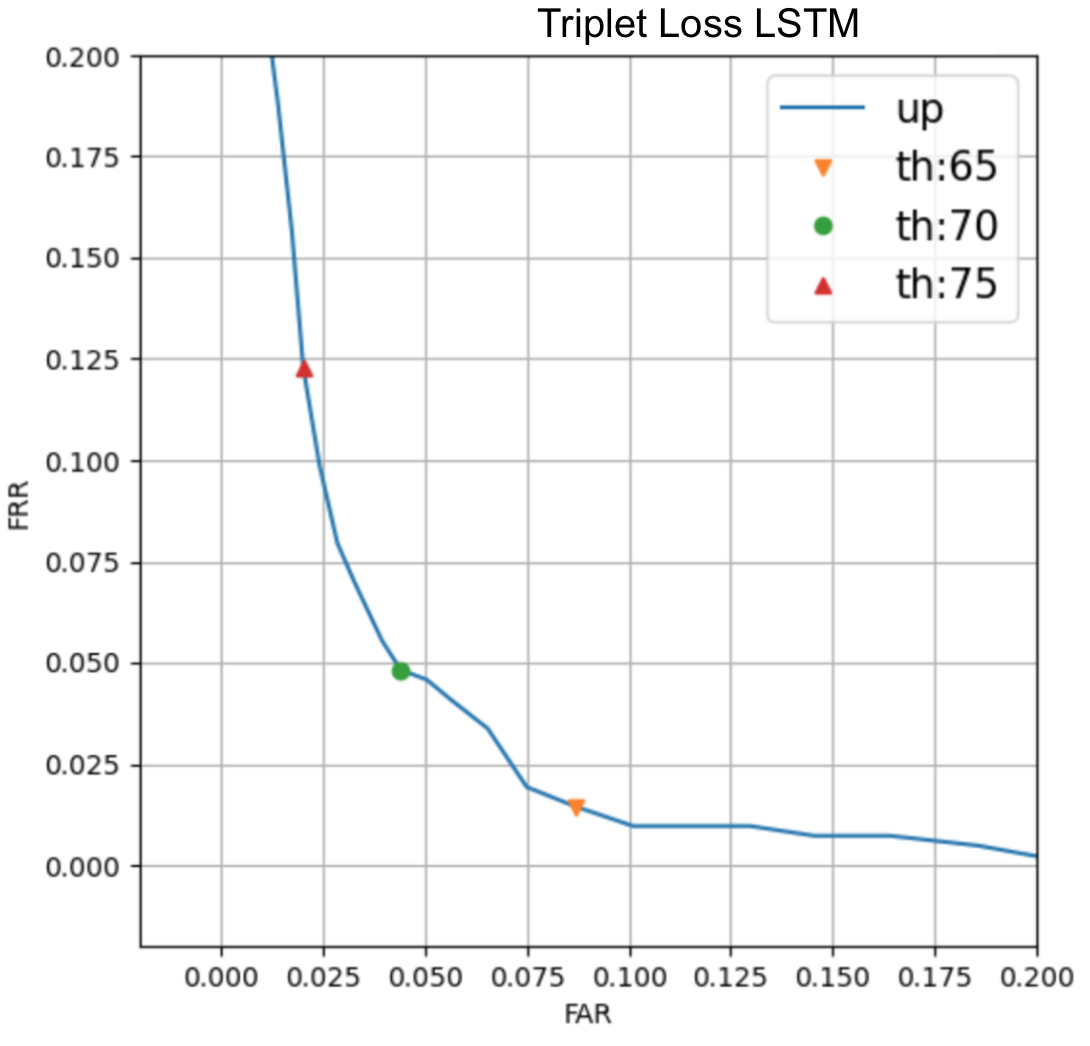}
	\caption{Left: Score histogram for an 247Kb LSTM model. Right: DET curve. The range of FAR/FRR is truncated for better visualization.}
	\label{fig:lstm_hist}
\end{figure}

The left part of Fig.~\ref{fig:lstm_hist} shows the cosine similarity histogram of the 247Kb LSTM model's evaluation result for the phrase ``up''. A perfect model would have a vertical line (threshold) to separate the true positive (red) and true negative (blue) examples. The overlapping parts are the errors, which, depending on the threshold, can be divided into False Accepts (FA) and False Rejects (FR). When sliding the threshold from 0 to 1 with a stride of 0.01, we plot the False Accept Rates (FARs) and False Reject Rates (FRRs) corresponding to each threshold in the DET curve on the right in Fig.~\ref{fig:lstm_hist}. We use the area under the DET curve (AUC) under the full FAR/FRR range to measure model performance independently of threshold. To measure model performance independently of enrollment phrases, we average the AUC across phrases. Lower values are better, and this is the default metric used in the rest of the paper to rank models and checkpoints. 

\section{Experiment Results }
\label{sec:exp_results}
Section.~\ref{ssec:tts_exps} explained three scenarios for exploring TTS data impact on improving custom KWS models. This section will show the experiment results and analyses.

\subsection{TTS Model Experiments with No Real Data}
\label{ssec:tts_no_real}

\begin{figure}[!ht]
	\centering
	\includegraphics[width=0.4\textwidth]{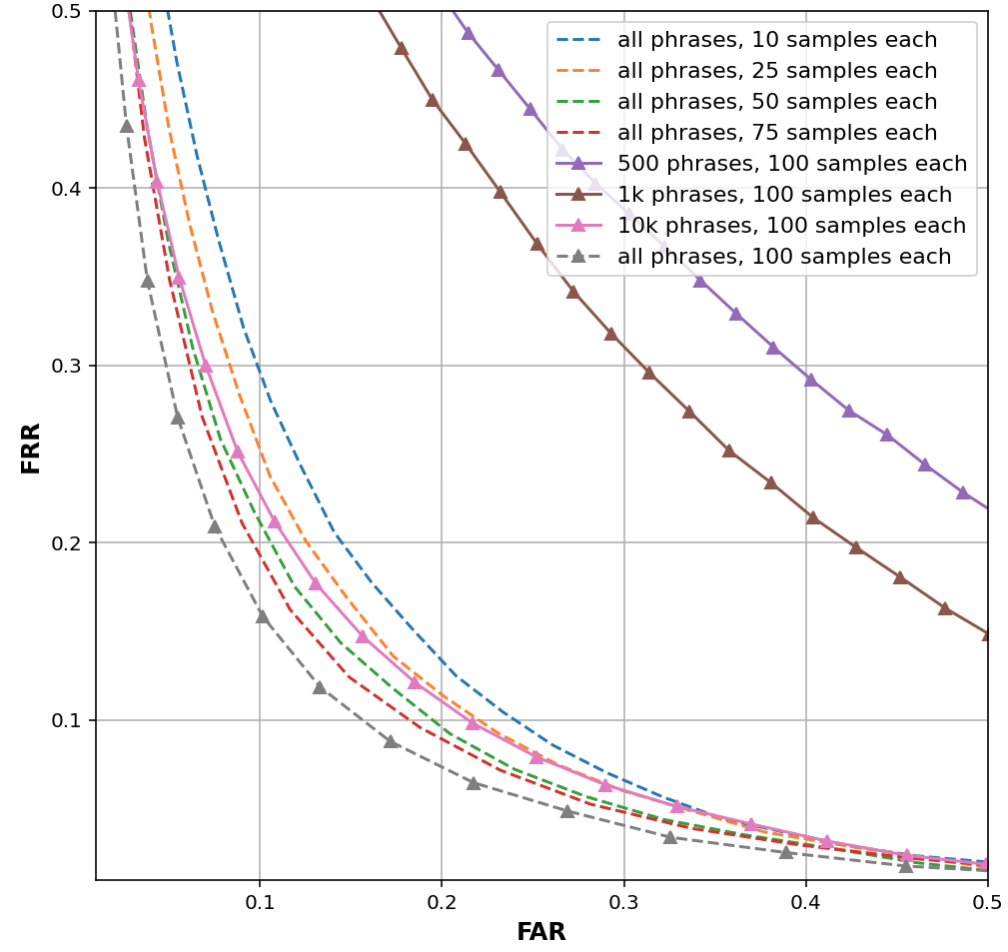}
	\caption{The ``No real data'' scenario. DET curves for models trained with varying phrase diversity and varying utterance samplings per phrase.}
	\label{fig:no_real_data_merge}
\end{figure}

We analyze the model improvements from TTS data by varying the amount of unique TTS phrases and varying amount of utterance samples per phrase. Table.~\ref{tab:no_real_data} top rows shows that when the number of unique phrases increases from 500 to all phrases: 38k (with 100 samples per phrase), the Equal Error Rates (EERs) and AUCs improve monotonically. The DET curves with triangles in Fig.~\ref{fig:no_real_data_merge} show that the improvement is consistent over different operating points. When fixing the phrase diversity at 38k and gradually increasing sampled utterance from 10 to 100, as shown in bottom rows in Table.~\ref{tab:no_real_data} and the dotted DET curves in Fig.~\ref{fig:no_real_data_merge}, model quality improves monotonically. In the environment without any real data, both TTS phrase diversity and sample sizes contribute proportionally to model performance.

\begin{table}[]
\centering
\resizebox{7cm}{!}{%
\begin{tabular}{l|l|l|l}
\begin{tabular}[c]{@{}l@{}}Amount of \\ unique phrases\end{tabular} & \begin{tabular}[c]{@{}l@{}}Sampled utterance\\ per phrase\end{tabular} & EER (\%) & AUC (\%) \\ \cline{1-4}         
500                                                                 & 100                                                                    & 34.42    & 28.90    \\
1k                                                                  & 100                                                                    & 30.17    & 24.20    \\
10k                                                                 & 100                                                                    & 15.09    & 7.68     \\
38k                                                                 & 100                                                                    & 12.62    & 5.93     \\ \cline{1-4}         
38k                                                                 & 10                                                                     & 16.88    & 9.32     \\ 
38k                                                                 & 25                                                                     & 15.67    & 8.34     \\
38k                                                                 & 50                                                                     & 14.59    & 7.22     \\
38k                                                                 & 75                                                                     & 13.78    & 6.99     \\
38k                                                                 & 100                                                                    & 12.62    & 5.93    

\end{tabular}}

\caption{No real data scenario. Model improvements by applying varying amounts of TTS phrases (Top 4 rows) and utterance samplings per phrase (Bottom 5 rows). Performance is measured by Equal Error Rate(EER) and Area under DET curve (AUC).}
\label{tab:no_real_data}
\end{table}

\subsection{TTS Model Experiments with Low Real Data}
\label{ssec:tts_low_real}
\begin{figure}[!ht]
	\centering
	\includegraphics[width=0.4\textwidth]{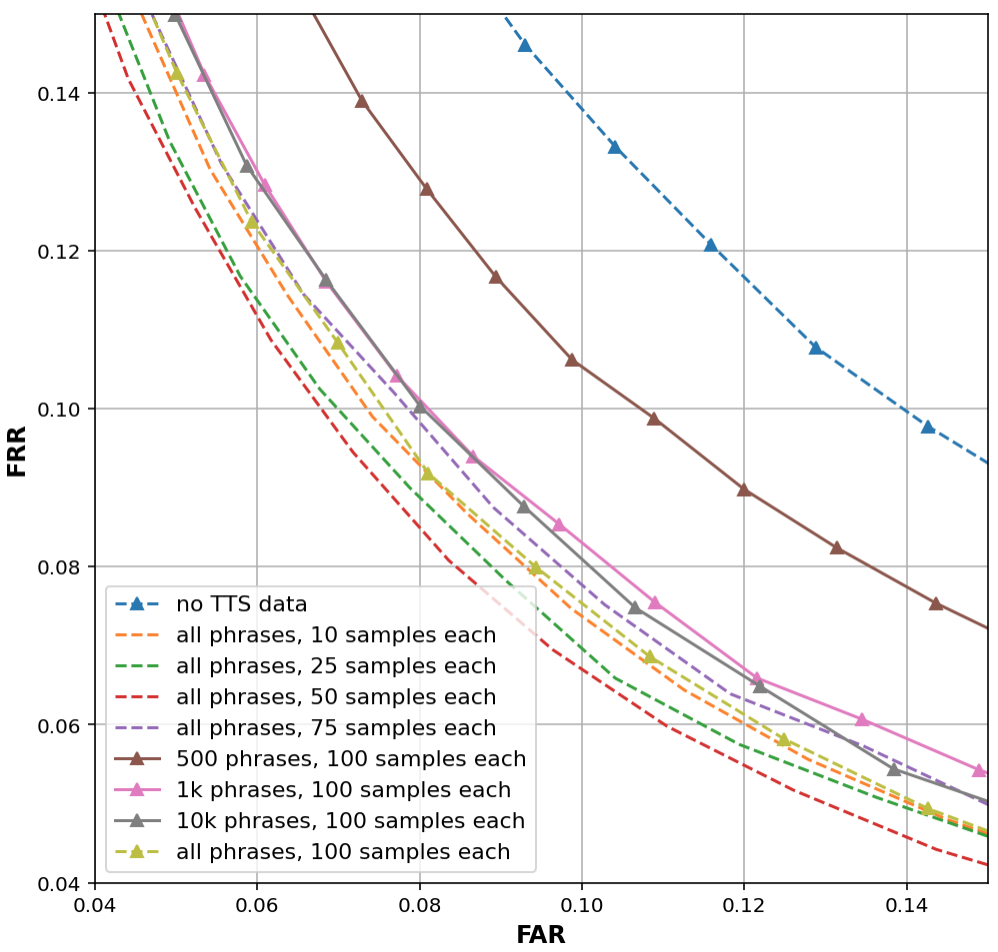}
	\caption{The ``Low real data scenario'' with 50k utterances. DET curves for models trained with varying phrase diversity and varying utterance samplings per phrase.}
	\label{fig:one_real_data_merge}
\end{figure}

In the scenario in which some real utterances are available but not in sufficient numbers for training, we explore the impact of TTS data by varying the number of unique phrases and utterance samples per phrase. 
In this experiment we use a model trained with 50k real utterances as a benchmark baseline. Shown in Table.~\ref{tab:one_real_data} and the DET curves with triangles in Fig~\ref{fig:one_real_data_merge}, as the number of unique TTS phrases increase, the EER improves from 11.73\% to 8.7\% (25.8\% rel.) for a model trained with 38k different phrases, and AUC improves from 5.15\% to 2.94\% (42.9\% rel.).
Moreover, with a fixed amount of 38k unique phrases, as the amount of utterance samples per phrase varies, the following trend occurs: the model continuously improves as the number of samples increases at the beginning. However, as shown in the bottom half of Table~\ref{tab:one_real_data}, and in the dotted curves in Fig.~\ref{fig:one_real_data_merge}, the improvement peaks when the number of utterances per phrase reaches 50 and declines after adding more examples. One explanation for this effect is that extra TTS data might overshadow the real data contribution so the model gets less information from real utterances. In the best case, when the model is trained with 38k phrases and 50 utterances per phrase, the model get further improved to 8.19\% EER (30.1\% rel. improvement) and AUC 2.64 (48.7\% rel. improvement).

\begin{table}[]
\centering
\resizebox{7cm}{!}{%
\begin{tabular}{l|l|l|l}
\begin{tabular}[c]{@{}l@{}}Amount of\\ unique phrases\end{tabular} & \begin{tabular}[c]{@{}l@{}}Number of utterances\\ per phrase\end{tabular} & EER (\%) & AUC (\%) \\ \cline{1-4}         
0                                                                  & 100                                                                    & 11.73                 & 5.15     \\
500                                                                & 100                                                                    & 10.17                 & 4.09     \\
1k                                                                 & 100                                                                    & 9.22                  & 3.17     \\
10k                                                                & 100                                                                    & 9.13                  & 3.07     \\
38k                                                                & 100                                                                    & 8.7                   & 2.94     \\ \cline{1-4}         
38k                                                                & 0                                                                      & 11.73                 & 5.15     \\ 
38k                                                                & 10                                                                     & 8.68                  & 2.92     \\
38k                                                                & 25                                                                     & 8.48                  & 2.85     \\
38k                                                                & 50                                                                     & 8.19                  & 2.64     \\
38k                                                                & 75                                                                     & 8.88                  & 2.91     \\
38k                                                                & 100                                                                    & 8.7                   & 2.94    
\end{tabular}}
\caption{``Low real data scenario'' with 50k real utterances as baseline. Model quality changes as a function of the number of TTS phrases (top 5 rows) and the number of utterances per phrase (bottom 6 rows). Performance is measured by Equal Error Rate (EER) and Area under DET curve (AUC).}
\label{tab:one_real_data}
\vspace{-5mm}
\end{table}

\subsection{TTS Model Experiments with Varying Real Data}
\label{ssec:how_much_real_data}

\begin{figure}[!ht]
	\centering
	\includegraphics[width=0.5\textwidth]{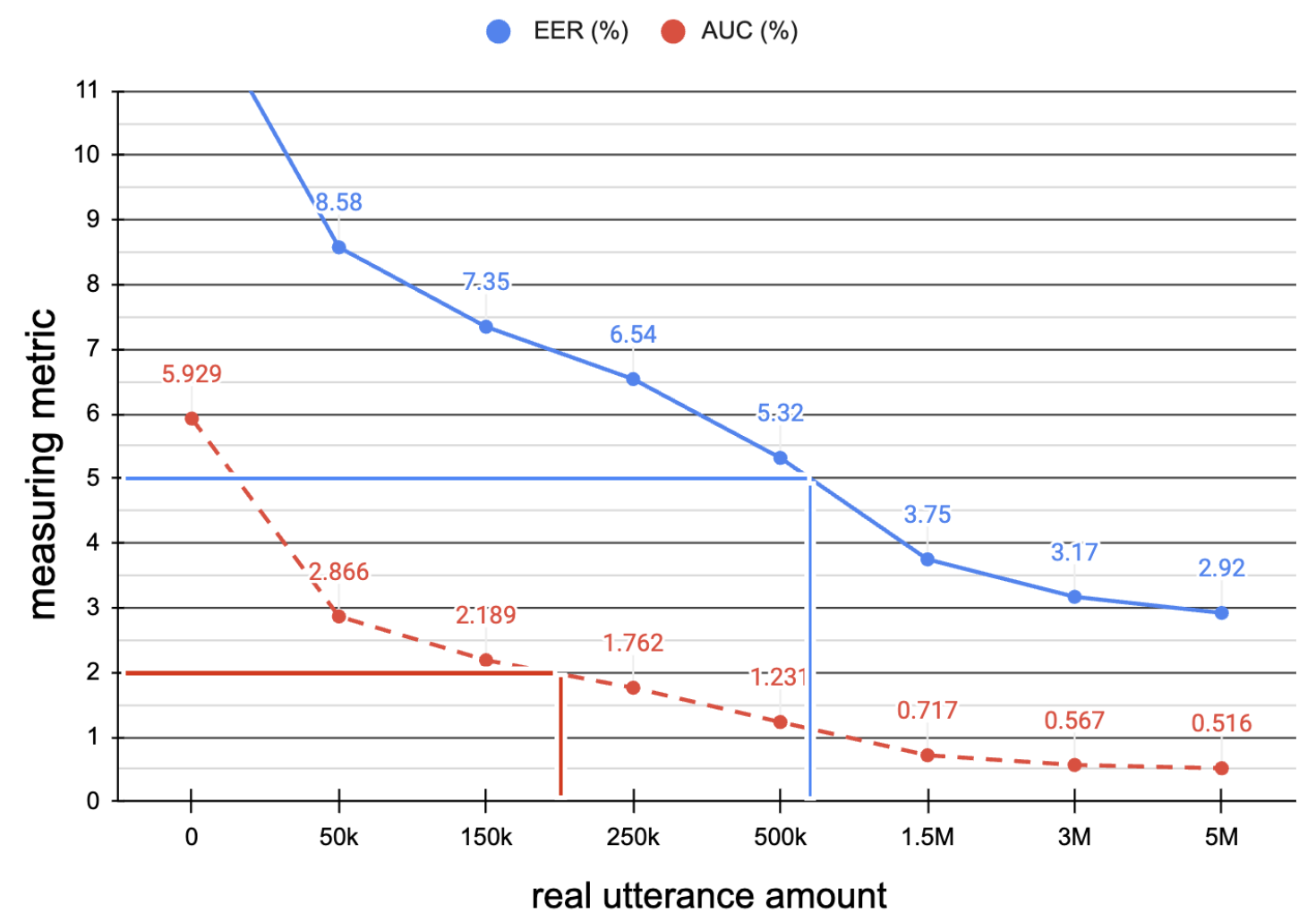}
	\caption{EER and AUC trends with increasing real utterances for model training. The TTS data with 38k phrases and 100 utterances per phrase are used in all experiments.}
	\label{fig:trend_vary_real_plot}
\end{figure}


We explored the impact of TTS data for model improvements in two scenarios: ``no real data'' and ``low real data''. Collecting real data is an expensive process but still necessary when the model is required to perform at high quality. To help resource planning and inform product decisions, we start with a model baseline that uses only TTS data (with 38k unique phrases and 100 utterances per phrase), and record the EER and AUC metrics when increasing the amount of the real training utterances to 0, 50k, 150k, 250k, 500k, 1M, 3M, and 5M.  We plot the EER and AUC trends in Fig.~\ref{fig:trend_vary_real_plot}.
It is easy to see that the amount of the real utterances is positively correlated model performance.
To find amount of real utterances needed to achieve a quality target, we interpolate the data points in Fig.~\ref{fig:trend_vary_real_plot}. For example, to achieve 5\% or lower EER, the interpolation shows about 700k real utterances need to be collected. Similarly, to achieve 2\% or lower AUC, we can see about 200k real utterances should be collected.

\section{Conclusion}
\label{sec:conclusion}
In this paper, we propose Synth4Kws --- a framework to leverage synthetic speech data for developing customizable keyword spotting models. With this framework, we carried out a systematic study on the impact of TTS data on user-defined keyword spotting tasks in various resource environments. In ``no real data'' experiments, we found that model quality monotonically improves when increasing TTS phrase diversity or utterances per phrase. In ``low real data'' experiments, with a baseline model with 50k real data, selecting the right amount of TTS phrases and utterance samples can improve EER by 30.1\% and AUC by 48.7\%. We also noticed that extra TTS data could overshadow the real data and produce worse results. Lastly, we interpolate the curve composed of model quality v.s. real data quantity and infer the amount of real data needed on top of TTS data to achieve different quality targets, which could be informative for resource planning. Finally although our experiment training and evaluations are based on the English language and one word utterances, we believe the results will generalize to other languages and multi-word keyword spotting tasks.



\clearpage
\bibliographystyle{IEEEtran}
\bibliography{mybib}

\end{document}